# Dynamics of Coupled Non-Identical Systems with Period-Doubling Cascade

A.P. Kuznetsov[1,2], I.R. Sataev[1], J.V. Sedova[1,2]

[1]*Institute of Radio-Engineering and Electronics of Russian Academy of Sciences,*

*Saratov Division, Saratov, Russia*

[2]*Saratov State University, Saratov, Russia*

## Abstract

Structure of bifurcation diagram in the plane of parameters controlling period-doublings for the system of coupled logistic maps is discussed. The analysis is carried out by computing the charts of dynamical regimes and charts of Lyapunov exponents giving showy and effective illustrations. The critical point of codimension two at the border of chaos is found. It is a terminal point for the Feigenbaum critical line. The bifurcation analysis in the vicinity of this point is presented.

## 1. Introduction

What can we know in advance about regularities of transition to chaos in coupled system, provided the individual dynamics of subsystems is well studied? The answer depends strongly on the subsystem dynamics type and character of coupling. We shall consider here the particular case of subsystems demonstrating period doubling transition to chaos (see [1-7]). The case of coupled *identical* subsystems is rather well investigated. For such systems the possible regimes of dynamics and scenarios of transition to chaos were determined [1-6]. There exists an alternative approach to research in the case of *non-identical* systems. Each of subsystems is characterized by its own control parameter responsible for period-doublings. We shall independently adjust each of these parameters. Then naturally a problem concerning the organization of the bifurcation diagram in the plane of these parameters may be posed. Similar approach has been advanced in a series of papers [8-12] concerning a situation when only one of subsystems influences another. The method turned out to be constructive and has leaded to discovery of new type of critical behavior when the subsystems are successively brought to the threshold of chaos by tuning the control parameters responsible for period-doublings. In the present paper we shall develop similar approach with regard to mutually coupled systems with symmetrical coupling. Some preliminary information can be found in [7] where the opportunity of existence of quasiperiodic regimes in the vicinity of diagonal on a plane of control parameters of subsystems was found out. With the help of charts of dynamical regimes and of Lyapunov exponent charts [8, 9, 13-15] we research in detail the structure of a plane of parameters responsible for period-doublings in subsystems for the system of two coupled logistic maps. Also the bifurcation analysis at the border of the domain of quasiperiodic dynamics will be given. This analysis reveals a critical point of codimension two [14-17] that is new for the given type of systems. It is an accumulation point for the sequence of corresponding codimension-2 bifurcation points and is a terminal point of the Feigenbaum critical line.

Bifurcation analysis in the section 3 was carried out with the help of program package CONTENT [18].

## 2. Global structure of bifurcation diagram in the parameters space

As object for analysis we choose system of coupled logistic maps in the next form:

$$\begin{cases} x_{n+1} = \lambda_1 - x_n^2 + \varepsilon(x_n - y_n), \\ y_{n+1} = \lambda_2 - y_n^2 + \varepsilon(y_n - x_n). \end{cases} \tag{1}$$

Here $x$, $y$ are the dynamical variables, $\lambda_1, \lambda_2$ are parameters responsible for period-doublings in subsystems, $\varepsilon$ is a parameter of coupling.

To visualize and examine the complex organization of parameters space we use three supplementing each other methods, namely a method of charts of dynamical regimes, a method of Lyapunov exponent charts and bifurcation analysis. We shall further give the necessary explanations.

The method of charts of the dynamical regimes consists in computing the domains of stability of different attractors via scanning the parameter space of the system under investigation. Different domains are shown in the chart with different colors (or, for example, gradations of gray tone) [8, 9, 13-15]. Such diagrams were computed for system (1) at various values of coupling parameter $\varepsilon$ (Fig. 1). The enlarged fragments of these charts are presented in the Fig. 2. The used color palette (legend) is shown underneath of Fig. 1. The white color corresponds to attractors, whose period was not determined (chaos, quasiperiodic regimes). The gray color corresponds to areas, where the orbit escapes to infinity (divergence). The largest areas of stability are indicated by numerals meaning the period of cycle.

It should be noted, that coupled maps systems, in general, are characterized by multistability. Therefore the structure of charts can depend on a choice of initial conditions and the way of scanning of parameters plane. We choose a method of scanning with memory, when the last point of the orbit at the previous node of the mesh in the parameters plane is used for setting initial conditions for iterations at the next node. It is the so-called technique of "inheritance" of initial conditions. By choosing the initial conditions at the first point of the scan line we may control and examine the structure of domains of stability for each of the coexisting attractors. In this paper we shall only discuss the structure of the one sheet which is chosen by fixing initial conditions at origin at the bottom edge of the scan area. Organization of other sheets is very similar to this particular case. Global picture would be discussed elsewhere.

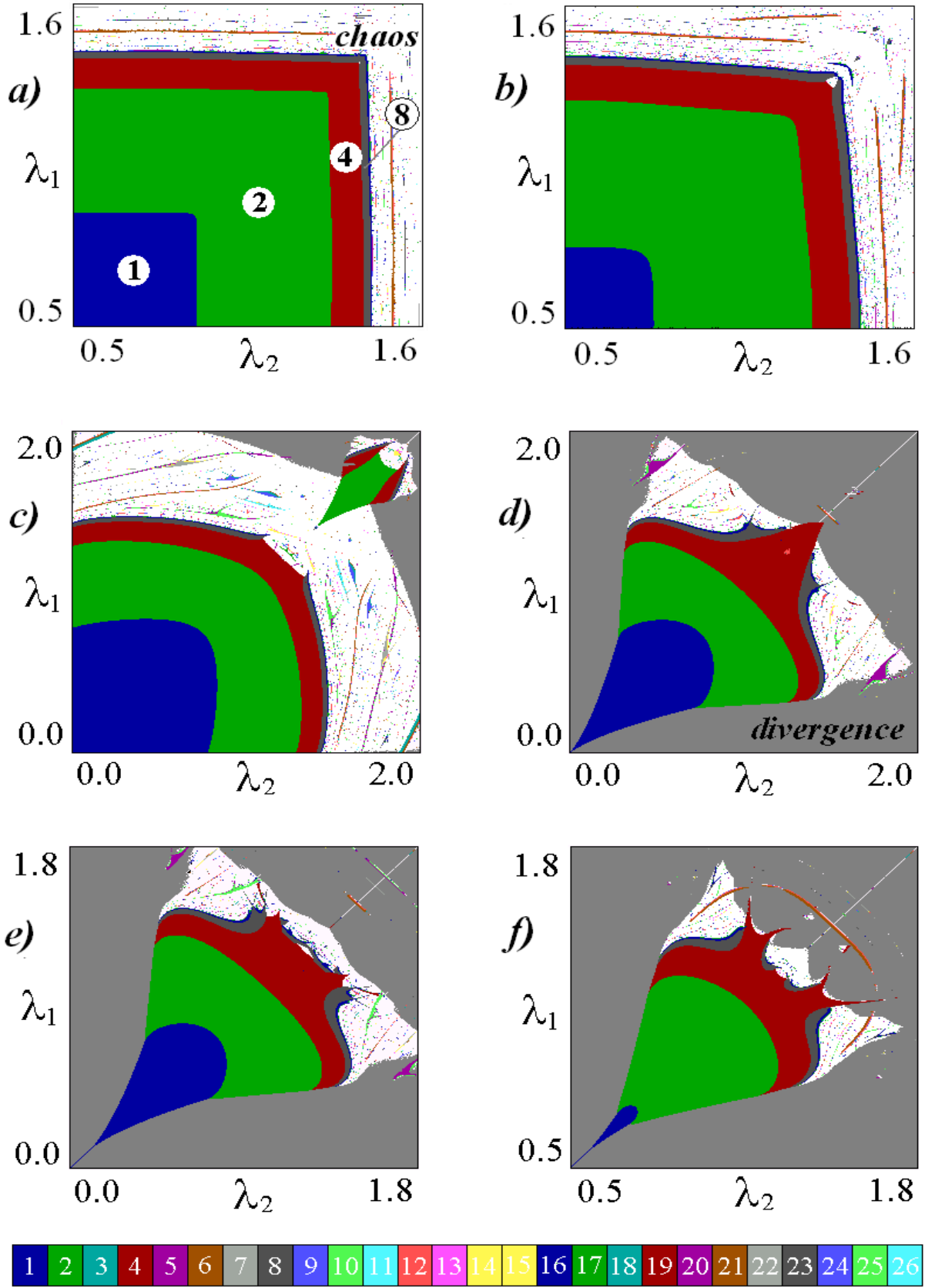

Figure 1. The charts of dynamical regimes for system of coupled non-identical logistic maps (1) on the plane $(\lambda_2, \lambda_1)$ at the different values of coupling parameter $\varepsilon$: $\varepsilon = 0.01$ (*a*), $\varepsilon = 0.04$ (*b*), $\varepsilon = 0.2$ (*c*), $\varepsilon = 0.5$ (*d*), $\varepsilon = 0.6$ (*e*), $\varepsilon = 0.9$ (*f*). The legend at the bottom specifies the correspondence between colors and cycle periods. Also for the best comprehension on the fragment (*a*) the largest areas of stability are indicated by numerals meaning the period of cycle.

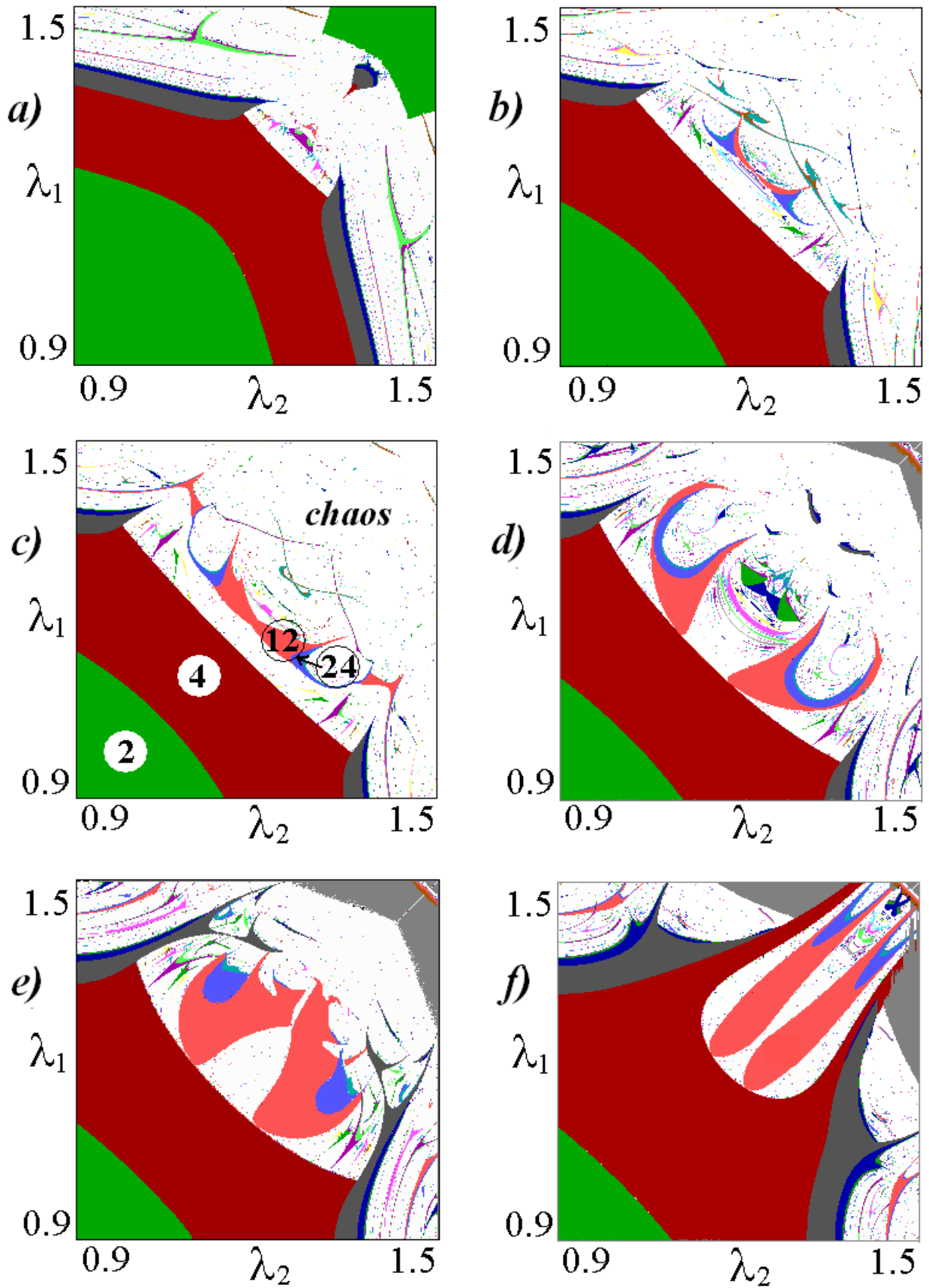

Figure 2. Transformation of synchronization tongues of the system (1) on the plane $(\lambda_2, \lambda_1)$ in the range of coupling parameter from ε=0.14 up to 0.49: ε=0.14 (*a*), ε=0.25 (*b*), ε=0.3 (*c*), ε=0.4 (*d*), ε=0.44 (*e*), ε=0.49 (*f*). On the fragment (*c*) the largest areas of stability are indicated by numerals meaning the period of cycle.

The charts in Figs.1,2 have some common features, in particular they are symmetric with respect to the main diagonal which corresponds to the case of identical subsystems. Far enough from the diagonal the Feigenbaum scenario of period-doublings takes place. Near the diagonal the domain of quasiperiodic regimes on the base of "period 4-cycle" with the immanent hierarchy of synchronization tongues is observed. In Fig.2 this region is shown under magnification allowing to consider its structure in more details at variation of coupling parameter from the value $\varepsilon = 0.14$ up to $\varepsilon = 0.49$. From Figures 1 and 2 we can see that above-mentioned area has rather complicated internal organization and undergoes considerable metamorphosis in process of growth of coupling parameter. From below the domain of quasiperiodic regimes is limited by the line of Neimark-Saker bifurcation (details see further) on the base of a regular regime with the period 4. On the right and on the left to this domain one can observe the lines of period-doublings accumulating at Feigenbaum critical curves. Visually, due to the presence of region of quasiperiodic dynamics Feigenbaum curves have a gap on charts. We shall show below that it is really so and these lines have specific terminal points.

Interesting feature of problem under consideration is the presence of synchronization tongues of two types. In Fig.2 (with the exception of Fig.2*f*), close to the left and right edges of quasiperiodic area we can see a very narrow synchronization tongues of high periods of the "classic" horn form, their spikes originate from the Neimark-Saker bifurcation curve. Nearby the diagonal it is possible to see another family of periodic regions similar to half-rounds, which also can be regarded as tongues but originating from the line of symmetry. These tongues are generated pairwise, and their pedestals are united on the diagonal of a chart. In Fig.3 one of the most typical tongues of this second family is shown enlarged. The synchronization tongue corresponds to the basic period 12, i.e. it is the triple period of a cycle on the base of which the Neimark-Saker bifurcation takes place. Typical phase portraits in various points of a parameters plane are represented in the insets. Portraits of attractors corresponding to the chaos, periodic and quasiperiodic regimes are displayed. It is necessary to remark that underneath the bottom edge of the tongue we observe quasiperiodic regimes (fragments 5 and 7). Corresponding phase portraits look like four closed curves because quasiperiodic regimes arise on the base of a period 4-cycle. Period-12 cycle can be interpreted as a resonance cycle on this invariant curve.

Inside the tongue with the basic period 12 (when moving from the area *A* to the area *B* in Fig.3) we can observe period-doubling bifurcation and then come to the domain with stable period-24 cycle. The peculiarity of Fig.3 is that following further inside the area with period 24 we shall come to the other area of period 12 (region *B*). Underneath the bottom edge of the tongue we observe chaotic dynamics instead of quasiperiodic one (see fragment 1). Moving this way still further we reveal a window of period 12, which lies in the chaotic region and stretches along the lines of period-doublings and the critical Feigenbaum line. Thus the foregoing tongue of the second family is conjugated to the periodic window from the supercritical area of Feigenbaum critical line.

Undertaken consideration of Fig.2 implies that synchronization tongues of the second type are typical objects in a parameters plane of the coupled systems with period-doublings.

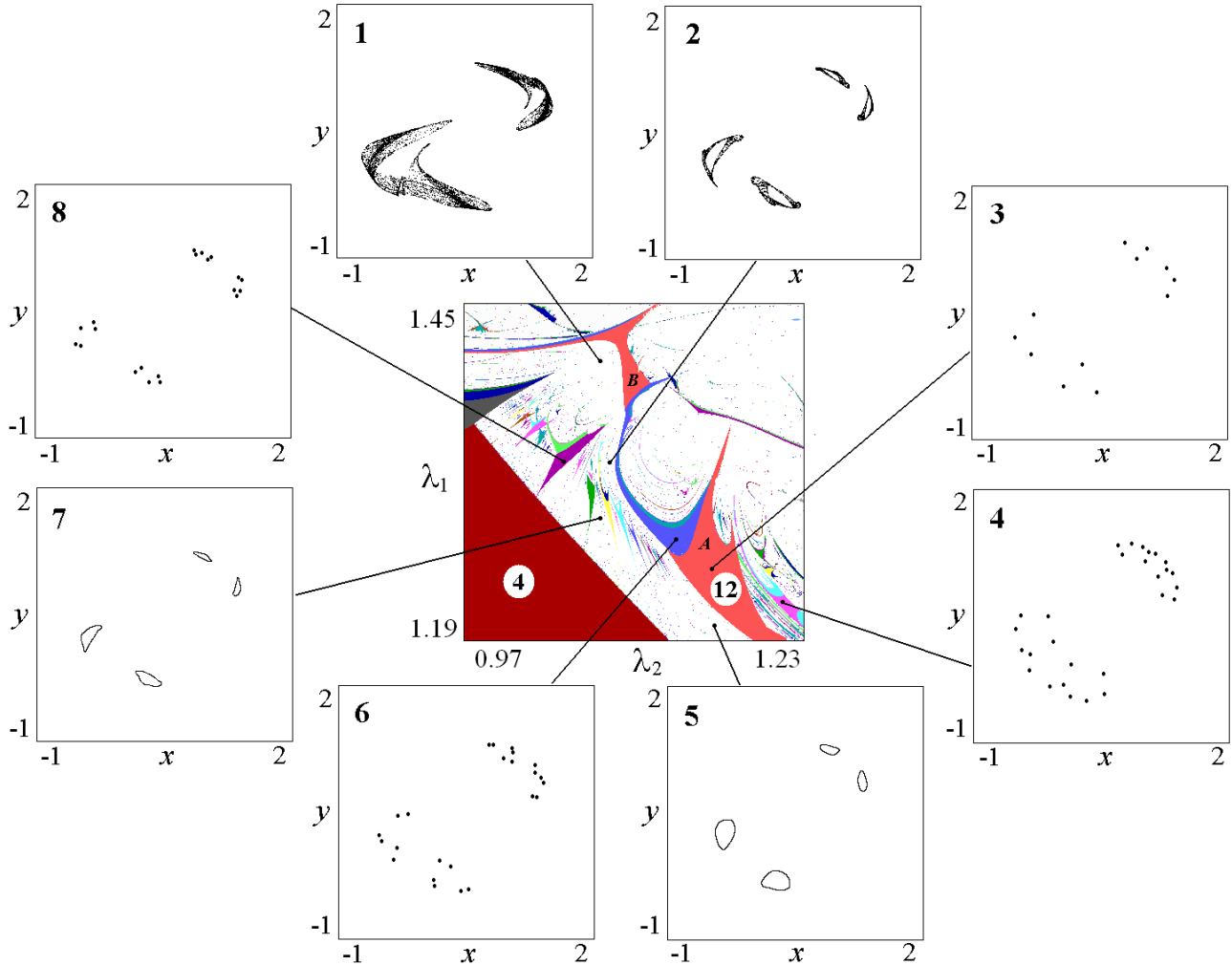

Figure 3. Portraits of attractors on the phase plane $(x, y)$ in the different points of the parameters plane $(\lambda_2, \lambda_1)$ for the value of coupling parameter $\varepsilon = 0.3$. In the center of Figure we present the fragment of the chart of dynamical regimes for $\varepsilon = 0.3$ that demonstrates the presence of tongues of the second family.

Further evolution of the domain of stability of period-12 cycle shows, that it eventually appears to be resonance 1:3 synchronization tongue of classical type (first family tongue according to our terminology) and the connection with supercritical Feigenbaum periodic window is lost (Fig. 2*d*). At the same time near the diagonal in Fig. *2d,e* it is possible to see the regions of "high periods-cycles" (tongues of the second type), still laying inside the quasiperiodic areas.

In Fig.4 phase portraits of attractors are shown at various points of parameters plane $(\lambda_2, \lambda_1)$ for the value of coupling parameter $\varepsilon = 0.4$ in a vicinity of synchronization tongue that originates from the Neimark-Saker bifurcation curve (first family).

Another method, which gives effective information about dynamics of system, is the method of Lyapunov exponent charts [19-22]. To compute such chart we calculate the value of Lyapunov exponent $\Lambda$ at each point of a mesh in parameters plane. Then its value is coded by the gradations of gray color according to the next rule. White color corresponds to the value of $\Lambda$ that is close to zero. Points with negative values of $\Lambda$ are coded by the gray shadings: the greater the absolute value of $\Lambda$ the darker is shading. Black color designates all positive values of Lyapunov exponent. Thus it is possible to distinguish quasiperiodic regime with zero Lyapunov exponent from chaos for which this index is positive. White color also indicates points at which iterative process is diverging. In Fig.5 we demonstrate the Lyapunov exponent charts for two non-identical logistic maps (1) on a plane of parameters controlling period-doublings in partial systems. Parameters values are the same as in Fig.2. These charts are convenient to visualize regions of quasiperiodic regimes: these areas looks like white areas with immanent entrainment horns inside.

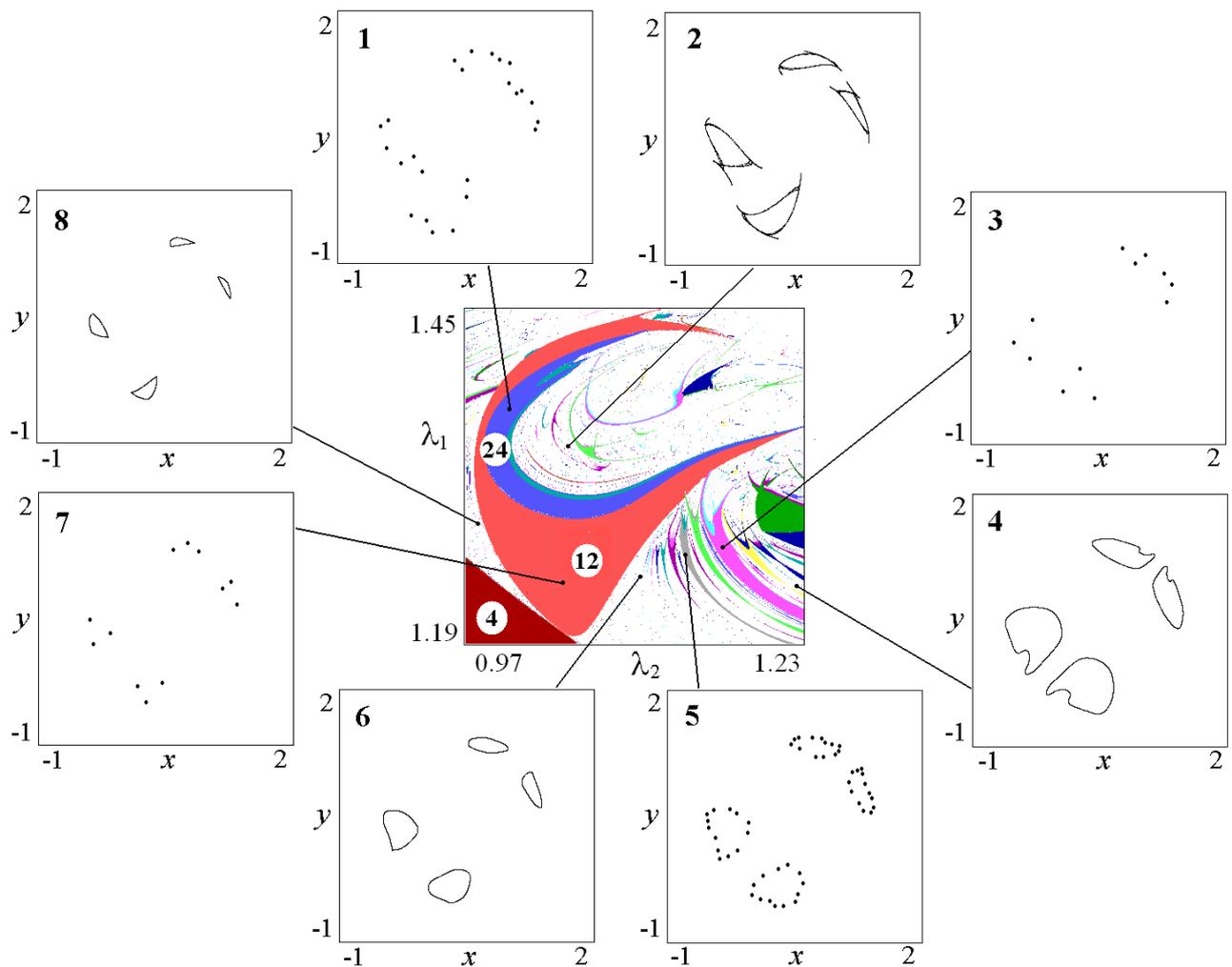

Figure 4. Portraits of attractors on the phase plane ($x$, $y$) at the different points of the parameters plane ($\lambda_2$, $\lambda_1$) for the value of coupling parameter $\varepsilon = 0.4$. In the center of Figure we present the fragment of chart of dynamical regimes for $\varepsilon = 0.4$ that demonstrated the presence of tongues of the first (classic) family.

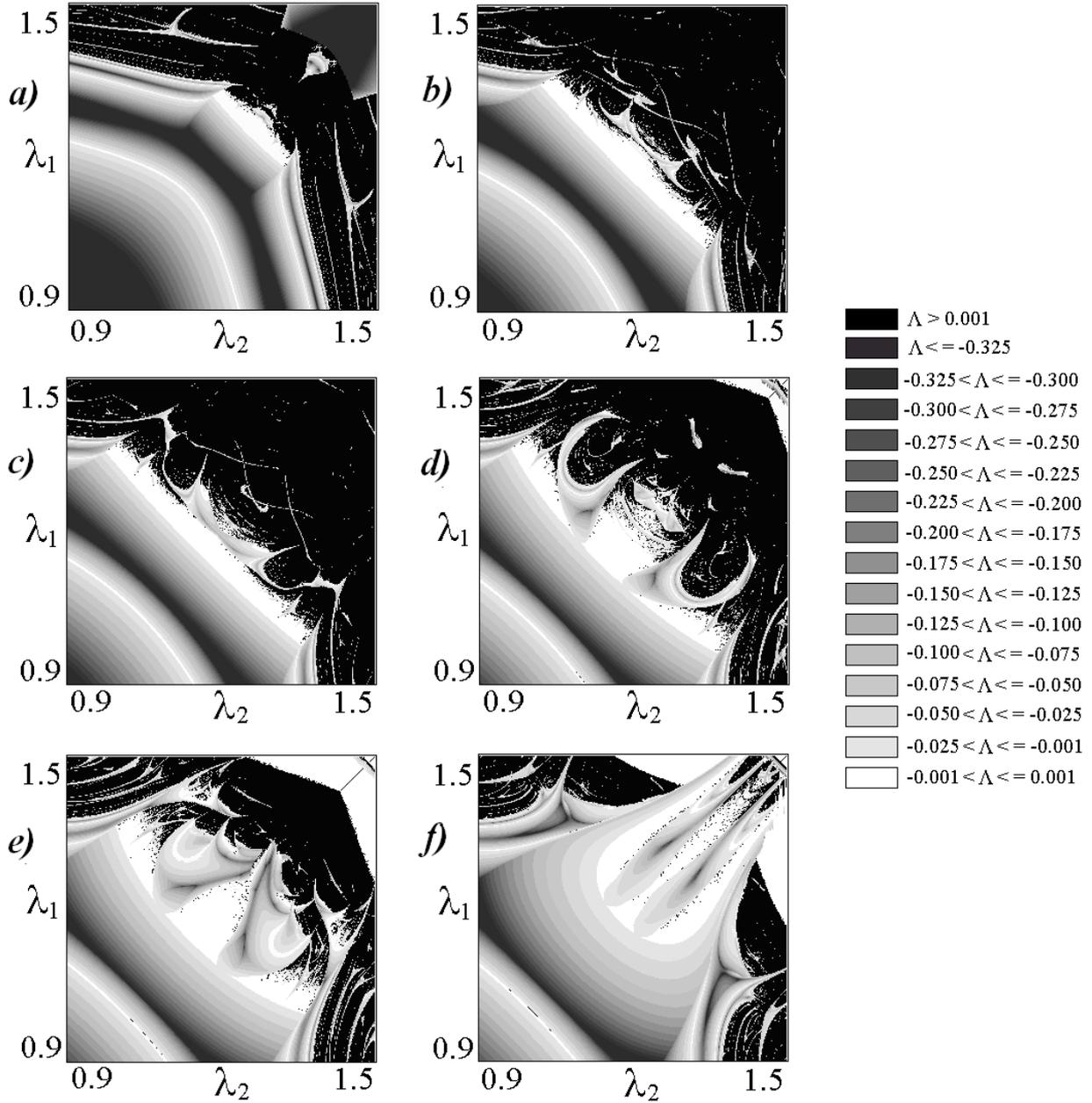

Figure 5. Lyapunov exponent charts for the system (1) in the plane $(\lambda_2, \lambda_1)$ in the range of changing of coupling parameter from ε=0.14 to 0.49: ε=0.14 (*a*), ε=0.25 (*b*), ε=0.3 (*c*), ε=0.4 (*d*), ε=0.44 (*e*), ε=0.49 (*f*). Parameters values are the same as in Fig.2. The legend is shown on the right.

## 3. Terminal points of Feigenbaum critical lines and bifurcation analysis in their neighborhood

Let us discuss the bifurcation diagram in more detail. In Fig.6*a* the chart of dynamical regimes is reproduced for the coupling parameter value *ε*=0.4, while in Fig.6*b* for the same parameter values the bifurcation curves and codimention-2 bifurcation points are shown which constitute the boundaries of stability domains presented in Fig.6*a*. One can easily see breakdown of the Feigenbaum period doubling cascade in the vicinity of the diagonal, where (as we already have shown) quasiperiodic regimes and synchronization tongues are observed. Beginning with a cycle of period 4 the corresponding lines of period-doubling bifurcation are terminated before reaching the diagonal at codimension-2 bifurcation points. For period-4 cycle it is a point of resonance 1:2 (the point $R_2$ in Figures, both multipliers of a cycle are equal to $\mu_{1,2} = -1$) and for cycles of higher periods there are *terminal* points at which multipliers are equal to $\mu_1 = -1$ и $\mu_2 = 1$ (in Figures these points are designated as *FF* (*Flip-Fold*)).

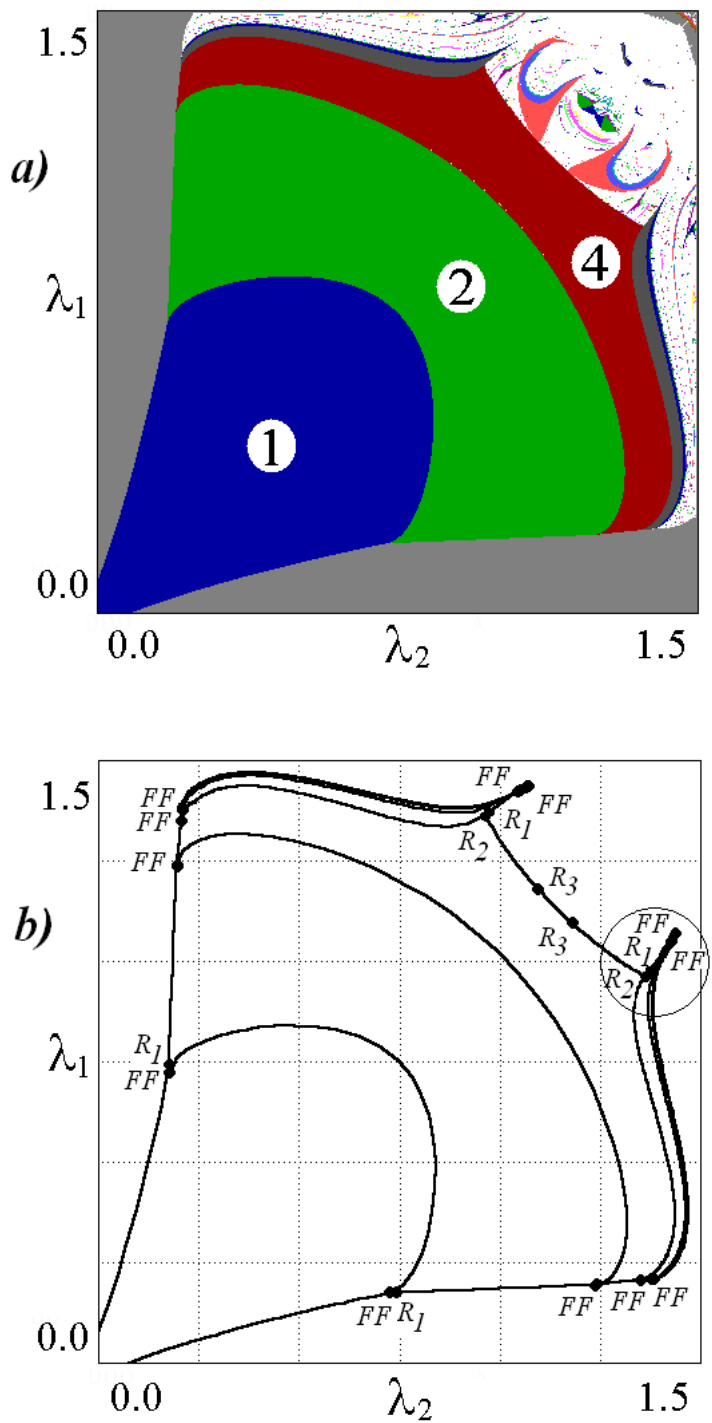

Figure 6. The chart of dynamical regimes (*a*) for the system (1) at $\varepsilon = 0.4$ and bifurcation diagram in the parameters plane ($\lambda_2$, $\lambda_1$) depicting basic bifurcation lines and points (*b*). *PD* is a point of period-doubling bifurcation, *FF* is a flip-fold terminal point, $R_1$ is a resonance 1:1 point, $R_2$ is a resonance 1:2 point, $R_3$ is a resonance 1:3 point. By circle on fragment (*b*) the area is shown that is displayed in Fig.7*a* in the enlarged scale.

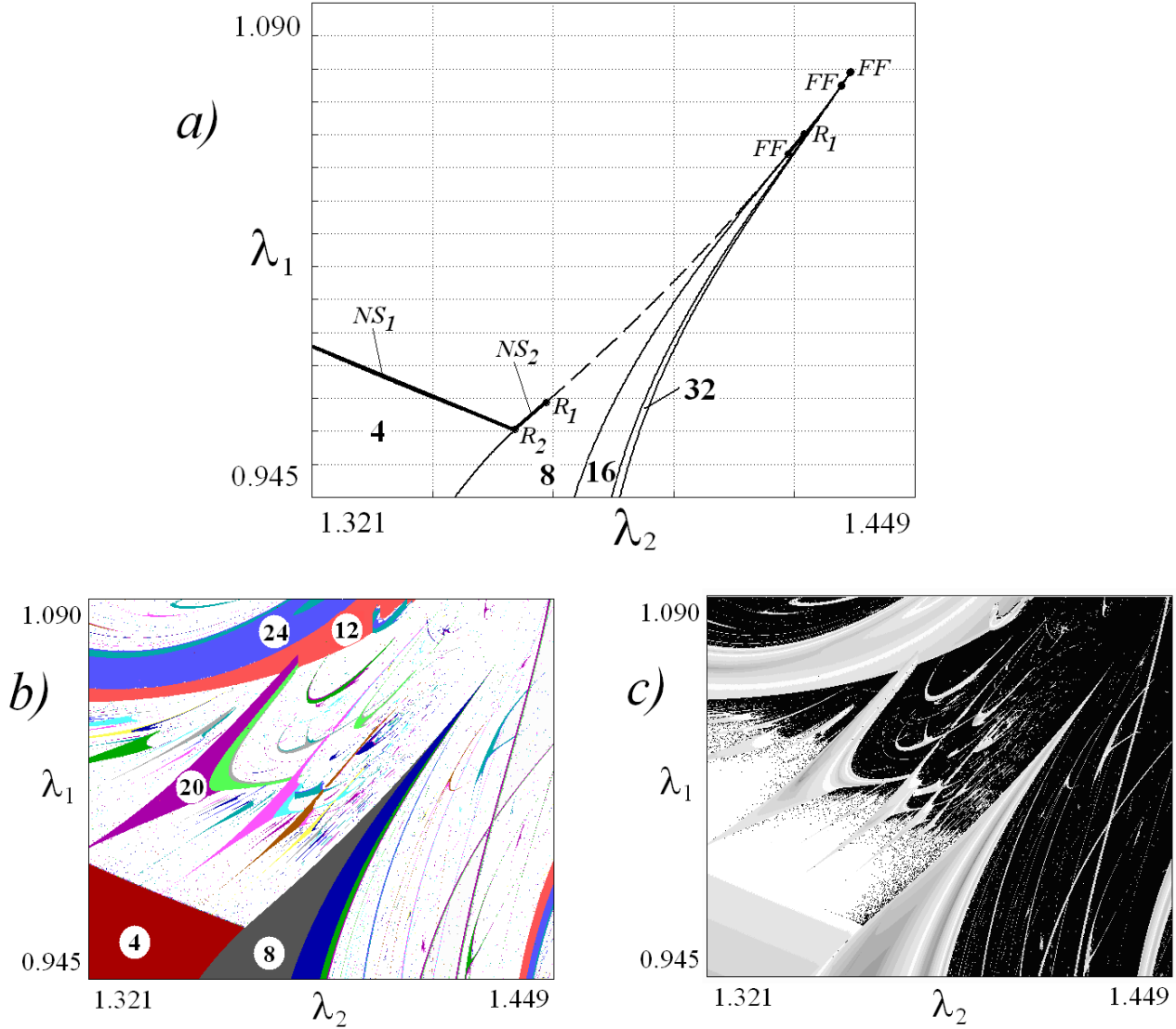

Figure 7. The picture of bifurcation lines in the vicinity of the codimention-2 critical point that is terminal for Feigenbaum critical curve (*a*). The regions of existence of stable cycles with periods 4, 8, 16 and 32 are marked by numerals; continuous lines corresponds to period-doubling bifurcation lines, dashed lines - to lines of rigid transition, thick lines – to lines of Neimark-Saker bifurcation *NS*. At the bottom we demonstrate corresponding fragments of chart of dynamical regimes (*b*) and Lyapunov exponent chart (*c*)

Bifurcation curves and points in the neighborhood of codimention-2 bifurcation point $R_2$ are shown in the enlarged scale in Fig.7*a*. In Fig.7*b,c* corresponding to Fig.7*a* chart of dynamical regimes and Lyapunov exponent chart are shown.

Emanating from the point $R_2$ is the curve of Neimark-Saker bifurcation $NS_1$ ($|\mu_{1,2}|=1$, it is marked by thick line), it is this curve which forms the edge of quasiperiodic area described earlier with the help of charts of dynamical regimes. It gives rise to Arnold synchronization tongues of various periods (see Fig.6*a*) including area of a resonance 1:3 ($R_3$ in Fig.6*b*). In the domain of doubled period cycle another Neimark-Saker bifurcation curve $NS_2$ originates from the $R_2$ point, it is also shown by thick line in Fig.7a. Thus the area of stability of period 8-cycle is limited from top (Fig.7a) partly by Neimark-Saker bifurcation line $NS_2$ and the fold bifurcation line ($\mu_1=1$, $|\mu_2|<1$, dashed line), they are separated by a codimension-2 bifurcation point $R_1$ (it is a resonance 1:1 bifurcation point, $\mu_{1,2}=+1$).

It should be noted, that a list of local bifurcations related to a resonance 1:2 point is exhausted by bifurcations demonstrated in Fig.7a. It is necessary to emphasize that $NS_2$ is the curve of *subcritical* Neimark-Saker bifurcation. In the vicinity of resonance there are still several lines of global bifurcations that are concerned with rearrangement of closed invariant manifolds arising as the results of bifurcations at lines $NS_1$ and $NS_2$ (see, for example, books [23, 24] where bifurcations in a vicinity of a resonance 1:2 are described in details).

In the Table 1 the coordinates for terminal points $FF$ of period-doubling bifurcation curves are presented. Apparently, this sequence accumulates to some limiting critical point. The similar sequence of $FF$-type bifurcation points was investigated in papers [14-17]. It was revealed that the critical point of this type is associated with period 2-cycle of doubling renormalization group transformation (or it may be regarded as a fixed point of quadrupling renormalization group transformation). This point is a terminal one for the Feigenbaum critical line, in other words it is a point at which the Feigenbaum line breaks. Such type of criticality was named *C-type*. It was shohwn, that at critical point of *C*-type the system demonstrates critical quasiattractor, infinite set of coexisting attractors, namely stable cycles of periods $2 \cdot 4^k$ and unstable cycles of periods $4^k$ ($k = 1, 2, \ldots, \infty$) [14-17].

**Table.** Terminal points $FF$ of system (1) for value of coupling parameter $\varepsilon = 0.4$

| Number of level | $X$ | $Y$ | $\lambda_2$ | $\lambda_1$ |
|---|---|---|---|---|
| 8 | 1.52748654438 | 0.832027050076 | 1.42211434872 | 1.04536723026 |
| 16 | 1.51924944092 | 0.925594316666 | 1.43342291359 | 1.06527796425 |
| 32 | 1.51549610195 | 0.944420633912 | 1.43540783724 | 1.06925872682 |
| 64 | 1.51462907469 | 0.947525094423 | 1.43560852092 | 1.06967100717 |
| 128 | 1.51448606773 | 0.948436165896 | 1.43574276220 | 1.06994764874 |
| 256 | 1.51444199033 | 0.948492315004 | 1.43572781554 | 1.06991681069 |
| 512 | 1.55015567623 | 0.691545135397 | 1.43574642151 | 1.06995521068 |
| 1024 | 1.55014968148 | 0.691536759962 | 1.43573942195 | 1.06994076322 |

It is necessary to note that at small values of parameter $\lambda_2$ at the another end of the Feigenbaum critical curve (and by virtue of symmetry at small values of parameter $\lambda_1$) we can also observe an accumulation of terminal points of period-doubling lines, however, in the given paper we study in detail only the area where domains of quasiperiodic dynamics and period-doublings converge.

## 4. Conclusions

The structure of bifurcation diagram in the control parameters space of the symmetrically coupled non-identical systems with period-doublings (namely logistic maps) is investigated. For such systems transition to chaos via breakup of the quasiperiodic dynamics and related phenomena are observed in the vicinity of the line of symmetry. There were revealed two families of synchronization tongues. Tongues of the first type have a "traditional" organization and originates from Neimark-Saker bifurcation curve. Tongues of the second family have no contact with this line and demonstrate themselves as symmetric pairs with spikes (or pedestals) incorporated on the line of symmetry in the control parameters plane. Region of quasiperiodic dynamics borders with the area, where period doubling scenario of transition to chaos is observed. Feigenbaum critical line terminates at the border of the area of quasiperiodicity. The terminal point is shown to be of C-type, which was found out earlier for model systems and is characterized by universal dynamics admitting renormalization group description. Therefore we expect that similar features of the organization of control parameters plane will be typical for other coupled discrete and continuous time systems demonstrating period-doubling bifurcations cascade.

*The work was supported by Russian Science Support Foundation, Ministry of Education and Science of Russian Federation and CRDF (grant CRDF BRHE REC-006 SR-006-X1/BF5M06 Y3-P-06-07), grant of President of Russian Federation (MK-4162.2006.2) and grant of Russian Fund of Basic Researches (06-02-16619).*

## References

1. V.S. Anishchenko, V.V. Astakhov, A.B. Neiman, T.E. Vadivasova, L. Schimansky-Geier, Nonlinear Dynamics of Chaotic and Stochastic Systems. Tutorial and Modern Development (Springer, Berlin, 2002).

2. E. Mosekilde, Y. Maistrenko, D. Postnov, Chaos Synchronization. Application to Living Systems (World Scientific Series on Nonlinear Science, 2002, Series A42).

3. C. Reick, E. Mosekilde, Physical Review **E52** (1995) 1418-1435.

4. J. Rasmussen, E. Mosekilde, C. Reick, Mathematics and Computers in Simulation **40** (1996) 247-270.

5. S. Yanchuk, Y. Maistrenko, E. Mosekilde, Physica **D154** (2001) 26-42.

6. S. Yanchuk, T. Kapitaniak, Physical Letters **A290** (2001) 139-144.

7. J.-M. Yuan, M. Tung, D.H. Feng, L.M. Narducci, Physical Review **A28** (1983) 1662-1666.

8. A.P. Kuznetsov, S.P. Kuznetsov, I.R. Sataev, International Journal of Bifurcation and Chaos **1** (1991) 839-848.

9. A.P. Kuznetsov, S.P. Kuznetsov, I.R. Sataev, International Journal of Bifurcation and Chaos **3** (1993) 139-152.

10. S.-Y. Kim, W. Lim, Y. Kim, Progress of Theoretical Physics **106** (2001) 17-37.

11. S.-Y. Kim, Physical Review **E59** (1999) 6585-6592.

12. S.-Y. Kim, W. Lim, Physical Review **E63** (2001) 036223.

13. S.P. Kuznetsov, Dynamical chaos (FizMatLit, Moscow, 2006). (In Russian)

14. A.P. Kuznetsov, S.P. Kuznetsov, I.R. Sataev, Physica **D109** (1997) 91-112.

15. S.P. Kuznetsov, A.P. Kuznetsov, I.R. Sataev, Journal of Statistical Physics **121** (2005) 697-748.

16. S.P. Kuznetsov, I.R. Sataev, Physica **D101** (1997) 249-269.

17. S.P. Kuznetsov, I.R. Sataev, Physical Review **E64** (2001) 046214.

18. Yu.A. Kuznetsov, V.V. Levitin, CONTENT: A multiplatform environment for continuation and bifurcation analysis of dynamical systems (Centrum voor Wiskunde en Informatica, Kruislaan 413, 1098 SJ Amsterdam, The Netherlands, 1997).

19. M. Marcus, B. Hess, Computers and Graphics **13** (1989) 553-558.

20. J. Rössler, M. Kiwi, B. Hess, M. Marcus, Physical Review **A39** (1989) 5954-5960.

21. M. Marcus, Computers in physics, September/October (1990) 481.

22. J.C. Bastos de Figueireido, C.P. Malta, International Journal of Bifurcation and Chaos **8** (1998) 281-293.

23. V.I. Arnold, Geometrical Methods in the Theory of Ordinary Differential Equations (Springer, 1988).

24. J. Guckenheimer, P. Holmes, Nonlinear Oscillations, Dynamical Systems, and Bifurcations of Vector Fields (Springer-Verlag, 1983).